\documentclass[english,pra,showpacs,showkeys,tightenlines,secnumarabic,11pt]{revtex4}
\usepackage[T1]{fontenc}
\usepackage[latin1]{inputenc}
\usepackage{amsmath}
\usepackage{graphicx}
\usepackage{amssymb}
\usepackage{epsfig}
\usepackage{graphics}
\usepackage[mathscr]{euscript}
\usepackage{psfrag}
\usepackage{pstricks}
\usepackage{pst-node}
\usepackage{babel}
\newcommand{\bbeta}{\mbox{\boldmath$\beta$}}
\newcommand{\esp}[1]{\, e^{\,\,\textstyle {#1}}}

\begin{document}
\title{Multi-parton correlations and "exclusive" cross sections}
\author{G. Calucci}
\email{giorgio.calucci@ts.infn.it}
\author{D. Treleani}
\email{daniele.treleani@ts.infn.it} \affiliation{ Dipartimento di
Fisica Teorica dell'Universit\`a di Trieste and INFN, Sezione di
Trieste,\\ Strada Costiera 11, Miramare-Grignano, I-34014 Trieste,
Italy.}
\begin{abstract}
In addition to the inclusive cross sections discussed within the
QCD-parton model, in the regime of multiple parton interactions,
different and more exclusive cross sections become experimentally
viable and may be suitably measured. Indeed, in its study of
double parton collisions, the quantity measured by CDF was an
"exclusive" rather than an inclusive cross section. The non
perturbative input to the "exclusive" cross sections is different
with respect to the non perturbative input of the inclusive cross
sections and involves correlation terms of the hadron structure
already at the level of single parton collisions. The matter is
discussed in details keeping explicitly into account the effects
of double and of triple parton collisions.
\end{abstract}



 \maketitle

\section{Introduction}

The growing importance of multiple parton interactions (MPI) at
high energy has stimulated a lot of interest in the phenomenon, in
view of the forthcoming results at the
LHC\cite{HERA-LHC}\cite{Perugia}. MPI are essential to describe
the features of the minimum bias and of the underlying
event\cite{Acosta:2004wqa}\cite{Acosta:2006bp}\cite{Sjostrand:1987su}\cite{Sjostrand:2004ef}\cite{Sjostrand:2006za}\cite{Butterworth:1996zw}\cite{Bahr:2008dy}
and may represent an important background in many channels of
interest at the LHC\cite{HERA-LHC}\cite{Perugia}, not only in
processes where the cross section is large, as the unitarity issue
at the origin of the effect might suggest, but also in cases where
the cross section is rather small, like for the search of the
Higgs boson\cite{DelFabbro:1999tf}\cite{Hussein:2007gj} or in the
case of the production of equal sign $W$ boson
pairs\cite{Kulesza:1999zh}\cite{Cattaruzza:2005nu}\cite{Perugia}.
On the other hand MPI are by themselves an interesting topic of
research, since by studying MPI one may obtain informations on the
multi-parton structure of the hadron\cite{HERA-LHC}\cite{Perugia}.

Up to now the direct observation of MPI has not been easy. The
direct measurement of MPI requires in fact the identification of
the final fragments of the multiple process and the reduction of
statistics, due to the request of large momentum exchange in each
hard interaction, has restricted considerably the possibilities of
a direct study of the phenomenon. To measure the MPI one needs
moreover to separate the background due to hard radiation. A given
multi-partons final state may in fact be produced either by a
multiple or by a single parton collision. The separation between
the two contributions has proven to be experimentally feasible in
the case of double parton
collisions\cite{Akesson:1986iv}\cite{Alitti:1991rd}\cite{Abe:1997bp}\cite{Abe:1997xk}.
The enhanced contribution of MPI at high energy  will facilitate
considerably direct studies of the phenomenon and one may
reasonably expect that the separation of the two different
contributions will be done more easily at the LHC, at least in the
simplest cases of MPI.

In the regime where MPI may be observed directly, in addition to the inclusive cross sections usually considered in large momentum exchange processes, one has the possibility to measure  diverse and more exclusive cross sections, computable in perturbation theory and linked differently to the hadron structure\cite{Calucci:2008jw}. In $pp$ interactions MPI are dominated by independent collisions, initiated by different pairs of partons\cite{Humpert:1984ay}\cite{Paver:1984ux}. A direct consequence is that the multi-parton inclusive cross sections are proportional to the moments of the distribution in the number of collisions\cite{Calucci:2008jw}. On the other hand, a statistical distribution may be characterized either by its moments or by its different terms. While the moments of the distribution in the number of collisions are measured by the multi-parton inclusive cross sections, the different terms of the distribution are measured by a different set of observables, which one may call "exclusive" cross sections. Inclusive and "exclusive" cross sections result from independent measurements and are linked in a different way to the hadron structure. The two sets of cross sections are however connected by sum rules.
By testing the sum rules, namely by looking at the number of terms needed to saturate the sum rules in a given phase space region, one measures the effects of unitarity corrections, which allows to control the consistency of the analysis and provides an additional handle to obtain information on the multi-parton correlations of the hadron structure.

Interestingly, in its study of MPI, the CDF experiment did not measure the inclusive cross section of
double parton scattering. The events selected where i fact only those which contained just double parton collisions, while all events with triple scatterings (about 17\% of the sample of all events with double parton scatterings) where removed\cite{Abe:1997xk}. The resulting quantity measured by CDF is hence different with respect to the inclusive cross sections usually discussed in large $p_t$ physics. In fact it represents precisely one of the "exclusive" cross sections
recently discussed\cite{Calucci:2008jw}.

While the inclusive cross sections are linked directly to the
multi-parton structure of the hadron the link of the "exclusive"
cross sections with the hadron structure is much more elaborate.
The requirement of having only events with a given number of hard
collisions implies that the corresponding cross section
(being proportional to the probability of not having any further
hard interaction) depends on the whole series of
multiple hard collisions. The number of hard partonic collisions which can be observed directly is nevertheless limited, which allows to discuss the "exclusive" cross sections
by expanding in the number of elementary interactions. Simple and directly testable connections between inclusive and "exclusive" cross
sections may hence be  established by saturating the sum rules, which link "inclusive" and exclusive cross sections, with a finite number of terms.

The purpose of the present paper is to discuss the
"exclusive" multiparton scattering cross sections, going up to the third order in the number of collisions and keeping two-body parton correlations explicitly into account, while the effects of the three-body parton
correlations in the hadron structure will be neglected.
Explicit expressions in terms of the two-body correlation parameters will
be derived in a few simplest cases, where correlations will be assumed to depend
only on the transverse coordinates.

\section{"Exclusive" cross sections}

In $pp$ collisions, the inclusive cross sections are basically the moments of the
distribution of the number of MPI\cite{Calucci:2008jw}. The most basic information on the distribution in the number of collisions, the average number, is hence given by the
single scattering inclusive cross section of the QCD parton model. Analogously the $K$-parton scattering inclusive cross
section $\sigma_K$gives the $K$th moment of the distribution in the number
of collisions and is related directly to the $K$-body parton
distribution of the hadron structure:

\begin{eqnarray}
\sigma_K&=&{1\over K!}\int D_A(x_1 \dots x_K;b_1 \dots
b_K)\hat{\sigma}(x_1x_1')\dots\hat{\sigma}(x_Kx_K')\nonumber\\&&\qquad\quad\qquad\times
D_B(x_1' \dots x_K';b_1-\beta \dots b_K-\beta)dx_1dx_1'd^2b_1\dots
dx_Kdx_K'd^2b_Kd^2\beta\nonumber\\&=&{\langle
N(N-1)\dots(N-K+1)\rangle\over K!}\sigma_{hard}\end{eqnarray}

\noindent where $\sigma_{hard}$ represents the contribution to the
total inelastic cross section due to all events with al least one
hard interaction, while $D(x_1 \dots x_K;b_1 \dots b_K)$ is the
$K$-partons density of the hadron structure, with transverse
parton coordinates $b_1 \dots b_K$ and fractional momenta $x_1
\dots x_K$,  $\beta$ is the hadronic impact parameter and
$\hat{\sigma}$ the parton-parton cross section.

A way alternative to the set of moments, to provide the whole
information of the distribution, is represented by the set of the
different terms of the probability distribution of multiple
collisions. Correspondingly, in addition to the set of the
inclusive cross sections $\sigma_K$, one may consider the set of the
"exclusive" cross sections $\tilde\sigma_N$, where one selects the events where $\it only$ $N$ collisions are present. One hence has:

\begin{eqnarray}
\sigma_{hard}\equiv\sum_{N=1}^{\infty}\tilde\sigma_N,\qquad\sigma_K\equiv\sum_{N=K}^{\infty}{N(N-1)\dots(N-K+1)\over
K!}\tilde\sigma_N
\end{eqnarray}

\noindent which represents also a set of sum rules connecting the inclusive and the "exclusive" cross sections.

While the non perturbative input to the inclusive cross section $\sigma_K$ is given by the $K$-parton distributions of the hadron structure, as implicit in Eq.(2), the non-perturbative input to the "exclusive" cross sections is given by an infinite set of multi-parton distributions. The request of being in a perturbative regime limits however the number of partonic collisions and the sum rules in Eq.(2) are saturated by a few terms, in such a way that the "exclusive" cross sections can be expressed by finite combinations of inclusive cross sections. A particular case is when all correlations
$C_n$ with $n>2$ are negligible (higher order correlation terms may be introduced in the picture of the interaction as sketched in the appendix). In that instance all sums in Eq.(2) can be performed\cite{Calucci:1991qq}\cite{Calucci:1997ii} and (neglecting all rescatterings) the hard cross section $\sigma_{hard}$ can be expressed in the following functional form:

\begin{eqnarray}\sigma_{hard}(\beta)=
  \Bigl[ 1&-&{\rm exp}\Bigl\{{-\int dudu'{\partial_J}\hat{\sigma}(u,u')
  {\partial_{J'}}}\Bigr\}\Bigr]{\cal Z}_A[J]{\cal Z}_B[J']\biggm|_{J=J'=1}\end{eqnarray}

\noindent where $u\equiv\{x,b\}$, $u'\equiv\{x',b'-\beta\}$ and
$\hat{\sigma}(u,u')$ is the interaction probability of the two
partons with coordinates $u$ and $u'$. For simplicity flavor
indices are omitted and the integration limits are set by the
limits of the phase space window where the MPI are observed. The
two-body parton correlations are introduced through the functional

\begin{eqnarray}
{\cal Z}[J+1]&\equiv&{\rm exp}\biggl\{\int D(u)J(u)du+{1\over
2}\int C(u,v)J(u)J(v)dudv\biggl\}\cr &=&\sum_{n}{1\over n!}\int
J(u_1)\dots J(u_n)D_n(u_1\dots u_n)du_1\dots du_n
\end{eqnarray}

\noindent  which generates the non-perturbative input of the $n$-parton inclusive cross sections, the inclusive $n$-body parton distributions $D_n(u_1\dots u_n)$. Under these conditions Eq.(3) can be worked out fully explicitly. One obtains\cite{Calucci:1991qq}\cite{Calucci:1997ii}

\begin{eqnarray}\sigma_{hard}(\beta)=1-{\rm exp}\Bigl[-{1\over 2}\sum_na_n-{1\over 2}\sum_nb_n/n\Bigr]
\end{eqnarray}

\noindent where

\begin{eqnarray}a_n=(-1)^{n+1}\int&& D_A(u_1)\hat{\sigma}(u_1,u_1')C_B(u_1',u_2')
       \hat{\sigma}(u_2',u_2)
       C_A(u_2,u_3)\dots\cr
       &&\qquad\qquad\qquad\qquad\cdots
       \hat{\sigma}(u_n,u_n')D_B(u_n')
  \prod_{i=1}^ndu_idu_i'\end{eqnarray}

\noindent and the chain, which starts with $A$, may end either with $A$ or with $B$, depending wether $n$ is odd or even. For $b_n$ one has

\begin{eqnarray}b_n=(-1)^{n+1}\int &&C_A(u_n,u_1)\hat{\sigma}(u_1,u_1')C_B(u_1',u_2')
           \dots\cr&&\qquad\qquad\qquad\cdots C_B(u_{n-1}',u_n')\hat{\sigma}(u_n',u_n)
       \prod_{i=1}^ndu_idu_i'\end{eqnarray}

\noindent and in this case only even values of $n$ are possible.

The exponential in Eq.(5) represents the probability of no interaction at a given impact parameter $\beta$. All "exclusive" cross sections can be obtained from the argument of the exponential. One may start from the partonic interaction probability

\begin{eqnarray}1-\prod_{i,j=1}^n(1-\hat\sigma_{ij})
\end{eqnarray}

\noindent where in the product each index assumes a given value
only once, in such a way that possible re-interactions are not
included. The probability of having only a single interaction is expressed by

\begin{eqnarray}\Biggl(-{\partial\over\partial g}\Biggr)\prod_{i,j=1}^n(1-g\hat\sigma_{ij})\Bigg|_{g=1}=\sum_{kl}\hat\sigma_{kl}\prod_{ij\neq kl}^n(1-g\hat\sigma_{ij})\Bigg|_{g=1}\end{eqnarray}

\noindent while the probabilities of a double and of a triple
collision are

\begin{eqnarray}{1\over 2!}\Biggl(-{\partial\over\partial g}\Biggr)^2\prod_{i,j=1}^n(1-g\hat\sigma_{ij})\Bigg|_{g=1}&=&{1\over 2!}\sum_{kl}\sum_{rs}\hat\sigma_{kl}\hat\sigma_{rs}\prod_{ij\neq
kl,rs}^n(1-g\hat\sigma_{ij})\Bigg|_{g=1}\nonumber\\
{1\over 3!}\Biggl(-{\partial\over\partial
g}\Biggr)^3\prod_{i,j=1}^n(1-g\hat\sigma_{ij})\Bigg|_{g=1}&=&{1\over
3!}\sum_{kl}\sum_{rs}\sum_{tu}\hat\sigma_{kl}\hat\sigma_{rs}\hat\sigma_{tu}\prod_{ij\neq
kl,rs,tu}^n(1-g\hat\sigma_{ij})\Bigg|_{g=1}\end{eqnarray}

\noindent and the corresponding expressions for the "exclusive"
cross sections

\begin{eqnarray}\Biggl(-{\partial\over\partial
g}\Biggr)\esp{-X(g)}\Bigg|_{g=1}&=&X'(g)\esp{-X(g)}\Bigg|_{g=1}\nonumber\\
{1\over2!}\Biggl(-{\partial\over\partial
g}\Biggr)^2\esp{-X(g)}\Bigg|_{g=1}&=&{1\over2!}\Bigl\{[X'(g)]^2-X''(g)\Bigr\}\esp{-X(g)}\Bigg|_{g=1}\nonumber\\
{1\over3!}\Biggl(-{\partial\over\partial
g}\Biggr)^3\esp{-X(g)}\Bigg|_{g=1}&=&{1\over3!}\Bigl\{X'''(g)+[X'(g)]^3-3X'(g)X''(g)\Bigr\}\esp{-X(g)}\Bigg|_{g=1}\nonumber\\
\end{eqnarray}

\noindent where $X={1\over2}(\sum a_n+\sum b_n/n)$.

It's convenient to expand $X$ and its derivatives in the
number of elementary collisions

\begin{eqnarray}
X&=&X_1+X_2+X_3+\dots\nonumber\\
\end{eqnarray}

\noindent  where

\begin{eqnarray}
X_1&=&\int D_A(u)\hat{\sigma}(u,u') D_B(u')dudu'\nonumber\\
X_2&=&-{1\over2}\Bigl[\int D_A(u_1)\hat{\sigma}(u_1,u_1') C_B(u_1',u_2')
       \hat{\sigma}(u_2',u_2)D_A(u_2)\prod_{i=1}^2du_idu_i'+A\leftrightarrow B\Bigr]\nonumber\\
&&\qquad-{1\over2}\int C_A(u_1,u_2)\hat{\sigma}(u_1,u_1') C_B(u_1',u_2')
       \hat{\sigma}(u_2',u_2)\prod_{i=1}^2du_idu_i'\nonumber\\
X_3&=&\int D_A(u_1)\hat{\sigma}(u_1,u_1') C_B(u_1',u_2')
       \hat{\sigma}(u_2',u_2)C_A(u_2,u_3)
       \hat{\sigma}(u_3,u_3')\nonumber\\
&&\qquad\times D_B(u_3')\prod_{i=1}^3du_idu_i'\nonumber\\
\end{eqnarray}

\noindent The derivatives at $g=1$ give

\begin{eqnarray}
X_1'(v,v')&=&D_A(v)\hat{\sigma}(v,v') D_B(v')\nonumber\\
X_2'(v,v')&=&-\Bigl[ D_A(v)\hat{\sigma}(v,v') \int C_B(v',u_1')
       \hat{\sigma}(u_1',u_1)D_A(u_1)du_1du_1'+A\leftrightarrow B\Bigr]\nonumber\\
&&\qquad-\int C_A(u_1,v)\hat{\sigma}(v,v') C_B(v',u_1')
       \hat{\sigma}(u_1',u_1)du_1du_1'\nonumber\\
X_3'(v,v')&=&\Bigl[ D_A(v)\hat{\sigma}(v,v')\int
C_B(v',u_1')
       \hat{\sigma}(u_1',u_1)C_A(u_1,u_2)
       \hat{\sigma}(u_2,u_2')\nonumber\\
&&\qquad\times D_B(u_2')\prod_{i=1}^2du_idu_i'\nonumber\\
&&\quad+\int D_A(u_1)\hat{\sigma}(u_1,u_1') C_B(u_1',v')
       \hat{\sigma}(v',v)C_A(v,u_2)
       \hat{\sigma}(u_2,u_2')\nonumber\\
&&\qquad\times D_B(u_2')\prod_{i=1}^2du_idu_i'\nonumber\\
&&\quad+\int D_A(u_1)\hat{\sigma}(u_1,u_1') C_B(u_1',u_2')
       \hat{\sigma}(u_2',u_2)C_A(u_2,v)
       \hat{\sigma}(v,v')\nonumber\\
&&\qquad\times D_B(v')\prod_{i=1}^2du_idu_i'\Bigr]\nonumber\\
\end{eqnarray}

\noindent and

\begin{eqnarray}
X_2''(v_1,v_1';v_2,v_2')&=&-\Bigl[ D_A(v_1)\hat{\sigma}(v_1,v_1') C_B(v_1',v_2')
       \hat{\sigma}(v_2',v_2)D_A(v_2)+A\leftrightarrow B\Bigr]\nonumber\\
&&\qquad- C_A(v_2,v_1)\hat{\sigma}(v_1,v_1') C_B(v_1',v_2')
       \hat{\sigma}(v_2',v_2)\nonumber\\
X_3''(v_1,v_1';v_2,v_2')&=&2\Bigl[
D_A(v_1)\hat{\sigma}(v_1,v_1') C_B(v_1',v_2')
       \hat{\sigma}(v_2',v_2)\int C_A(v_2,u_1)
       \hat{\sigma}(u_1,u_1')\nonumber\\
&&\qquad\times D_B(u_1')du_1du_1'\nonumber\\
&&\quad+ D_A(v_1)\hat{\sigma}(v_1,v_1')\int C_B(v_1',u_1')
       \hat{\sigma}(u_1',u_1)C_A(u_1,v_2)du_1du_1'\nonumber\\
       &&\qquad\times\hat{\sigma}(v_2,v_2')
 D_B(v_2')\nonumber\\
&&\quad+ \int D_A(u_1)\hat{\sigma}(u_1,u_1') C_B(u_1',v_1')du_1du_1'
       \hat{\sigma}(v_1',v_1)C_A(v_1,v_2)\nonumber\\
       &&\qquad\times\hat{\sigma}(v_2,v_2')
 D_B(v_2')\Bigr]\nonumber\\
X_3'''(v_1,v_1';v_2,v_2';v_3,v_3')&=&
6D_A(v_1)\hat{\sigma}(v_1,v_1') C_B(v_1',v_2')
       \hat{\sigma}(v_2',v_2) C_A(v_2,v_3)
       \hat{\sigma}(v_3,v_3') D_B(v_3')\nonumber\\
\end{eqnarray}

\noindent By substituting the expansions in the number of
elementary collisions in the expressions of the interaction
probabilities and by expanding the exponential, one obtains the expressions:

\begin{eqnarray}
\tilde\sigma_1'&=&(X'_1+X'_2+X'_3)(1-X_1-X_2+X_1\cdot X_1/2)\nonumber\\
2\times\tilde\sigma_2''&=&(X'_1\cdot X'_1+2X'_1\cdot X'_2-X_2''-X_3'')(1-X_1)\nonumber\\
3\times\tilde\sigma_3'''&=&{1\over2}(X'''_3+X_1'\cdot X_1'\cdot X_1'-3X_1'\cdot X_2'')\nonumber\\
\end{eqnarray}

\noindent where $\tilde\sigma_1'$ etc. are the "exclusive" cross sections, differentiated according with (14) and (15). The integrated "exclusive" cross sections hence are

\begin{eqnarray}
\tilde\sigma_1&=&X_1-X_1^2-X_1X_2+X_1^3/2+2X_2-2X_2X_1+3X_3\nonumber\\
2\times\tilde\sigma_2&=&X_1^2+4X_1X_2-2X_2-6X_3-X_1^3+2X_1X_2\nonumber\\
3\times\tilde\sigma_3&=&3X_3+(X_1)^3/2-3X_1X_2\nonumber\\
\end{eqnarray}

\noindent The sum rules of Eq.(2) are satisfied as follows

\begin{eqnarray}
\tilde\sigma_1+2\times\tilde\sigma_2+3\times\tilde\sigma_3=&&X_1\nonumber\\&&-X_1^2+2X_2+X_1^2-2X_2-X_1X_2+X_1^3/2-2X_2X_1+3X_3\nonumber\\
&&+4X_1X_2-6X_3-X_1^3+2X_1X_2+3X_3+(X_1)^3/2-3X_1X_2\nonumber\\
=&&X_1\equiv\sigma_S
\nonumber\\
\nonumber\\
2\times\tilde\sigma_2+6\times\tilde\sigma_3
=&&X_1^2-2X_2\nonumber\\
&&+4X_1X_2-6X_3-X_1^3+2X_1X_2+6X_3+(X_1)^3-6X_1X_2\nonumber\\
=&&X_1^2-2X_2\equiv2\times\sigma_D
\nonumber\\
\nonumber\\
6\times\tilde\sigma_3=&&6X_3+(X_1)^3-6X_1X_2
\equiv3!\times\sigma_T
\end{eqnarray}

\noindent where $\sigma_S$, $\sigma_D$ and $\sigma_T$ are respectively the single, double and triple parton scattering inclusive cross sections. Explicitly

\begin{eqnarray}
\sigma_S&=&X_1=\int D_A\hat{\sigma}D_B\nonumber\\
\sigma_D&=&{1\over2}[X_1^2-2X_2]={1\over2}\Bigl[\int D_A\hat{\sigma}D_B\cdot D_A\hat{\sigma}D_B+\int D_A\hat{\sigma}C_B\hat{\sigma}D_A\nonumber\\
&&\qquad\qquad\qquad\qquad\qquad+\int D_B\hat{\sigma}C_A\hat{\sigma}D_B+\int C_A\hat{\sigma}C_B\hat{\sigma}\Bigr]\nonumber\\
&&\qquad\qquad\quad={1\over2}\int[D_AD_A+C_A]\hat{\sigma}\hat{\sigma}[D_BD_B+C_B]
\end{eqnarray}

\noindent where $[DD+C]\equiv D_2$, the two body parton
distribution as defined in Eq.(4). An analogous expression may be
written for $\sigma_T$.

The relations (18) may be inverted

\begin{eqnarray}
\tilde\sigma_1&=&\sigma_S-2\sigma_D+3\sigma_T\nonumber\\
\tilde\sigma_2&=&\sigma_D-3\sigma_T\nonumber\\
\tilde\sigma_3&=&\sigma_T
\end{eqnarray}

\noindent which allow to express the scale parameters characterizing the double and triple parton collisions in terms of the single scattering inclusive cross section $\sigma_S$ and of the single and double parton "exclusive" cross sections $\tilde\sigma_1$ and $\tilde\sigma_2$:

\begin{eqnarray}
\sigma_D&=&\sigma_S-\tilde\sigma_1-\tilde\sigma_2={1\over2}{\sigma_S^2\over\sigma_{eff}}\nonumber\\
\sigma_T&=&{1\over3}(\sigma_S-\tilde\sigma_1-2\tilde\sigma_2)={1\over6}\sigma_S^3{1\over\tau\sigma_{eff}^2}
\end{eqnarray}

\noindent where the scale factor of the triple parton scattering cross section has been characterized by the dimensionless parameter $\tau$.

\section{Correlations in transverse space}

Multiple parton collisions are most important in the region of
small fractional momenta, where the large population of partons
may dilute correlations and e.g. correlations due to energy
conservation may not be of major importance. A simplest
possibility is hence to neglect altogether correlations in
fractional momenta and to work out some case where only
correlations in transverse space are present and which allows an
analytic treatment.  To disentangle the effect of correlations in
transverse space from other sources of correlation, we will
consider the instance where all correlation terms give zero, when
integrated on the transverse coordinates, in such a way that all
other variables remain uncorrelated and, in particular, the
distribution in the number of partons is Poissonian.

The actual dependence of the correlation on the transverse
variables is not prescribed by general principles, in this section
two choices are presented, a Gaussian shape and an exponential
shape, both were used in a previous
discussion\cite{Calucci:2008jw}.

\subsection{Gaussian density}

Using Gaussian distributions for the parton
densities in transverse space and for the correlations one obtains closed analytic expressions:

\begin{eqnarray}
D(x,b)&=&G(x)f(b)\nonumber\\
f(b)&=&g(b,R^2)\nonumber\\
C(x_1,x_2;b_1,b_2)&=&G(x_1)G(x_2)h(b_1,b_2)\nonumber\\
h(b_1,b_2)&=&c\cdot g(B,R^2/2)\bar h(b,\lambda^2)\nonumber\\
\bar h(b,\lambda^2)&=&{d\over d\gamma}\bar
g(b,\lambda^2/\gamma)\Big|_{\gamma=1}
\end{eqnarray}

\noindent where $G(x)$ represents the usual one-body parton
distribution, $\bar g(b,\lambda^2/\gamma)\equiv \eta\cdot
g(b,\lambda^2/\gamma)$,

\begin{eqnarray}
g(b,R^2)={1\over\pi R^2}{\rm exp}(-b^2/R^2)\nonumber
\end{eqnarray}

\noindent and

\begin{eqnarray}
{\bf B}&=&[{\bf b}_1+{\bf b}_2]/2\nonumber\\
{\bf b}&=&[{\bf b}_1-{\bf b}_2]
\end{eqnarray}

\noindent in such a way that the following relations hold

\begin{eqnarray}
&&\int d^2b g(b,R^2)=1,\qquad\int d^2b_2 g({\bf b}_1-{\bf b}_2,R_1^2)g({\bf b}_2,R_2^2)=g({\bf b}_1,R_1^2+R_2^2)
\nonumber\\
&&\int h(b_1,b_2)d^2b=0
\end{eqnarray}

\noindent One may define the correlation length $r_c$ as the value
of $b$ where the correlation term $\bar h(b,\lambda^2)$ changes
sign. With our definition of $\bar h(b,\lambda^2)$ one has
$r_c=\lambda$.

\noindent To define unambiguously the "correlation strength" $c$,
one needs to normalize properly the correlation term $\bar
h(b,\lambda^2)$ (which integrates to zero). Our choice is

\begin{eqnarray}
\int_{|b|\le r_c} \bar h(b,\lambda^2)d^2b=1\nonumber
\end{eqnarray}

\noindent which gives $\eta=e$, where $e$ the Euler's number.

\noindent The integrations on the transverse variables of the
terms in Eq.'s 13-15 give

\begin{eqnarray}
D_A\hat{\sigma}D_B&\rightarrow&\int  d^2b d^2\beta g({\bf b}-{\bbeta},R_A^2)g({\bf b},R_B^2)=1\nonumber\\
D_A\hat{\sigma}D_B\cdot D_A\hat{\sigma}D_B&\rightarrow&\int  d^2b_1d^2b_2 d^2\beta g({\bf b}_1-{\bbeta},R_A^2)g({\bf b}_1,R_B^2) g({\bf b}_2-{\bf \bbeta},R_B^2)g({\bf b}_2,R_B^2)\nonumber\\
&&={1\over2\pi (R_A^2+R_B^2)}\nonumber\\
D_A\hat{\sigma}D_B\cdot D_A\hat{\sigma}D_B\cdot D_A\hat{\sigma}D_B&\rightarrow&\int  d^2b_1d^2b_2d^2b_3 d^2\beta g({\bf b}_1-{\bbeta},R_A^2)g({\bf b}_1,R_B^2) g({\bf b}_2-{\bf \bbeta},R_A^2)\nonumber\\
&&\qquad\qquad\qquad\times g({\bf b}_2,R_B^2)g({\bf b}_3-{\bf \bbeta},R_A^2)g({\bf b}_3,R_B^2)\nonumber\\
&&={1\over3\pi^2 (R_A^2+R_B^2)^2}\nonumber\\
D_A\hat{\sigma}C_B\hat{\sigma}D_A
&\rightarrow&\int  d^2b_1d^2b_2 d^2\beta g({\bf b}_1-{\bbeta},R_A^2)h_B({\bf b}_1,{\bf b}_2)g({\bf b}_2-{\bbeta},R_A^2)\nonumber\\
&&={c_Be\over\pi}{\lambda_B^2\over (2R_A^2+\lambda_B^2)^2}\nonumber\\
D_A\hat{\sigma}C_B\hat{\sigma}D_A\cdot D_A\hat{\sigma}D_B
&\rightarrow&\int  d^2b_1d^2b_2 d^2b_3d^2\beta g({\bf b}_1-{\bbeta},R_A^2)h_B({\bf b}_1,{\bf b}_2)g({\bf b}_2-{\bbeta},R_A^2)\nonumber\\
&&\qquad\qquad\qquad\times g({\bf b}_3-{\bbeta},R_A^2)g({\bf b}_3,R_B^2)\nonumber\\
&&={c_Be\over3\pi^2}{2\lambda_B^2\over (R_A^2+R_B^2)(2R_A^2+\lambda_B^2)^2}\nonumber\\
C_A\hat{\sigma}C_B\hat{\sigma}&\rightarrow&\int  d^2b_1d^2b_2 d^2\beta h_A({\bf b}_1-{\bbeta},{\bf b}_2-{\bbeta})h_B({\bf b}_1,{\bf b}_2)\nonumber\\
&&={c_Ac_Be^2\over\pi}{2\lambda_A^2\lambda_B^2\over(\lambda_A^2+\lambda_B^2)^3}\nonumber\\
C_A\hat{\sigma}C_B\hat{\sigma}\cdot D_A\hat{\sigma}D_B&\rightarrow&\int  d^2b_1d^2b_2d^2b_3 d^2\beta h_A({\bf b}_1-{\bbeta},{\bf b}_2-{\bbeta})h_B({\bf b}_1,{\bf b}_2)\nonumber\\
&&\qquad\qquad\qquad\times g({\bf b}_3-{\bbeta},R_A^2)g({\bf b}_3,R_B^2)\nonumber\\
&&={c_Ac_Be^2\over3\pi^2}{4\lambda_A^2\lambda_B^2\over(R_A^2+R_B^2)(\lambda_A^2+\lambda_B^2)^3}\nonumber\\
D_A\hat{\sigma}C_B\hat{\sigma}C_A\hat{\sigma}D_B&\rightarrow&\int  d^2b_1d^2b_2d^2b_3 d^2\beta g({\bf b}_1,R_A^2)h_B({\bf b}_1-{\bbeta},{\bf b}_2-{\bbeta})h_A({\bf b}_2,{\bf b}_3)\nonumber\\
&&\qquad\qquad\qquad\times g({\bf b}_3-{\bbeta},R_B^2)\nonumber\\
&&={16c_Ac_Be^2\over3\pi^2R_A^2R_B^2}E(s_A^2,s_B^2,r^2)
\end{eqnarray}

\noindent where $s_{A,B}=(\lambda/R)_{A,B}$, $r=R_A/R_B$ and

\begin{eqnarray}
E\bigl(s_A^2,s_B^2,r^2\bigr)\equiv{s_A^2s_B^2\Bigl[32r^2+100+32r^{-2}+6s_B^2(1+4r^{-2})
+3s_A^2\bigr(3s_B^2+2(4r^2+1)\bigr)\Bigr]\over
\Bigr\{s_A^2\bigl[3s_B^2+2(4r^2+1)\bigr]+2\bigl[6+s_B^2
(1+4r^{-2})\bigr]\Bigr\}^3}\nonumber
\end{eqnarray}

\noindent Using Eq.s(20) and (21), all inclusive and "exclusive"
cross sections, up to the triple order in the number of parton
collisions, are expressed in terms of the single scattering
inclusive cross section $\sigma_S$, of the "effective" cross
section $\sigma_{eff}$ and of the parameter $\tau$. In the case of
collisions of two identical hadrons the explicit expressions of
$\sigma_{eff}$ and $\tau$ are

\begin{eqnarray}
{1\over\sigma_{eff}}={3\over8\pi \bar R^2}\Bigl\{1+c\cdot e
{16\times3\bar s^2\over(4+3\bar s^2)^2}+c^2\cdot e^2{2\over3\bar
s^2}\Bigr\}
\end{eqnarray}

\noindent and

\begin{eqnarray}
{1\over\tau\sigma_{eff}^2}={3\over16\pi^2 \bar R^4}\Bigl\{1+c\cdot
e{16\times 9\bar s^2\over(4+3\bar s^2)^2}+c^2\cdot
e^2\Bigl[{2\over3\bar s^2}+6\times64E\Bigl({3\bar
s^2\over2},{3\bar s^2\over2},1\Bigr)\Bigl]\Bigr\}
\end{eqnarray}

\noindent where $\bar s\equiv \lambda/\bar R$ ($\lambda=r_c$ is the correlation length) and $\bar
R^2\equiv{3\over2}R^2$ is the square hadron radius measured in the
generalized parton
distributions\cite{Mueller:1998fv}\cite{Belitsky:2005qn}.

\subsection{Exponential density}

 The case where the parton density has an exponential shape is better
 displayed in Fourier-transform representation:

\begin{eqnarray}
g(b,R^2)&=&\frac {1}{(2\pi)^2}\int e^{-i{\bf k\cdot b}}\frac
{1}{(1+k^2R^2)^2}d^2k \end{eqnarray}

\noindent  while all other terms defined in Eq(22) are redefined
according with this unique change. In particular, the correlation
term is

\begin{eqnarray}
h(b_1,b_2)&=&c\cdot g(B,R^2/2)\bar h(b,\lambda^2)\nonumber\\
\bar h(b,\lambda^2)&=&{d\over d\gamma}\bar
g(b,\lambda^2/\gamma)\Big|_{\gamma=1}=\frac {2}{(2\pi)^2}\int
e^{-i{\bf k\cdot b}}\frac
{(k\lambda)^2}{(1+k^2\lambda^2)^3}d^2k\nonumber
\end{eqnarray}

\noindent and $\bar g(b,\lambda^2/\gamma)\equiv\eta\cdot
g(b,\lambda^2/\gamma)$. As in the previous case, one defines the
correlation length $r_c$ as the value of the distance $|{\bf
b}_1-{\bf b}_2|$ where the correlation $\bar h(b,\lambda^2)$
changes sign. The relation with the parameter $\lambda$ is
$r_c=x_0\lambda$ with $x_0\simeq2.387$\cite{Calucci:2008jw}. The
normalization parameter $\eta$, defined by the requirement

\begin{eqnarray}
\int_{|b|\le r_c} \bar h(b,\lambda^2)d^2b=1\nonumber
\end{eqnarray}

\noindent is now $\eta\simeq3.456$.

The integrations on the transverse variables cannot be always
displayed in closed form, a relevant simplification is got by
taking the parameters $R$ and $\lambda$ to be equal in A and B.

\noindent In this case the integrations on the transverse
variables of the terms in Eq.'s 13-15 give

\begin{eqnarray}
D_A\hat{\sigma}D_B&\rightarrow&\int  d^2b d^2\beta g({\bf b}-{\bbeta})g({\bf b})=1\nonumber\\
D_A\hat{\sigma}D_B\cdot D_A\hat{\sigma}D_B&\rightarrow&\int  d^2b_1d^2b_2 d^2\beta g({\bf b}_1-{\bbeta})g({\bf b}_1) g({\bf b}_2-{\bf \beta})g({\bf b}_2)\nonumber\\
&&={1\over4\pi R^2}{1\over7}\nonumber\\
D_A\hat{\sigma}C_B\hat{\sigma}D_A&\rightarrow&\int  d^2b_1d^2b_2 d^2\beta g({\bf b}_1-{\bbeta}) h({\bf b}_1,{\bf b}_2)g({\bf b}_2-{\bbeta})\nonumber\\
&&={c\over4\pi R^2}\eta F\Bigl[{1\over s^2}\Bigr]\nonumber\\
F(a)&\equiv&{3+44a-36a^2-12a^3+a^4+12a(2+3a){\rm ln}(a)\over3(a-1)^6a}\nonumber\\
C_A\hat{\sigma}C_B\hat{\sigma}&\rightarrow&\int  d^2b_1d^2b_2 d^2\beta  h({\bf b}_1-{\bbeta},{\bf b}_2-{\bbeta}) h({\bf b}_1,{\bf b}_2)\nonumber\\
&&={c^2\over4\pi R^2}{2\eta^2\over15s^2}\nonumber\\
D_A\hat{\sigma}D_B\cdot D_A\hat{\sigma}D_B\cdot
D_A\hat{\sigma}D_B&\rightarrow&
\frac{1}{(4\pi)^2}\frac{1}{R^4} \int_0^{\infty}\frac{(1+x+y)[(1+x+y)^2+6xy]dx\;dy}{(1+x)^4 (1+y)^4[(1+x+y)^2-4xy]^{7/2}}\nonumber\\
&&\simeq {1\over(4\pi R^2)^2}\times 0.030 \nonumber\\\nonumber\\
D_A\hat{\sigma}D_B\cdot D_A\hat{\sigma}C_B\hat{\sigma}D_A
&\rightarrow&c\frac{\eta}{(4\pi)^2}\frac{1}{R^4}\int_0^{\infty}\!\!2\frac{(1+x/4+y)^2-xy/2}{(1+x/2+y)^3[(1+x+2y+(x/2-y)^2]^{3/2}}\nonumber\\&&\qquad\qquad\qquad\times\frac{s^2 y}{(1+s^2 y)^3} \frac{dx\;dy}{(1+x)^4(1+x/2)^2}\nonumber\\
&&\equiv {c\over(4\pi R^2)^2}{\eta} H\Bigr[{1\over s^2}\Bigl]
\nonumber\\
C_A\hat{\sigma}C_B\hat{\sigma}\cdot D_A\hat{\sigma}D_B&\rightarrow&c^2\eta^2\frac{1}{4\pi^2}\frac{1}{R^4}\frac{1}{s^2} \int_0^{\infty}\frac {dx}{(1+x)^4(1+x/2)^4}\frac{y^2\,dy}{(1+y)^6}\nonumber\\
&&\simeq {c^2\over(4\pi R^2)^2}{\eta^2\over s^2}\times 0.0256
\nonumber\\\nonumber\\
D_A\hat{\sigma}C_B\hat{\sigma}C_A\hat{\sigma}D_B&\rightarrow&c^2\eta^2\frac{1}{(4\pi
R^2)^2}\int_0^{\pi}4\frac {d\phi}{\pi}
\int_0^{\infty}\frac{dx}{(1+x)^2(1+x/2)^2}\frac{dy}{(1+y)^2(1+y/2)^2}
\nonumber\\
&&\qquad\qquad\qquad\times\frac{w}{(1+w)^3}\frac{w'}{(1+w')^3}
\nonumber\\&&\equiv {c^2\over(4\pi R^2)^2}\eta^2L\Bigr[{1\over
s^2}\Bigl]
\end{eqnarray}
\noindent where $s=\lambda/R$, $w=s^2\,[x+y/4-\sqrt{xy}\cos\phi]$
and $ w'=s^2\,[x/4+y-\sqrt{xy}\cos\phi]$.

The functions $F$, $H$ and $L$ defined above, are given in Table I
for four different values of the parameter $s^2.$

\begin{table}
\caption{The functions $F$, $H$ and $L$ in Eq.'s(29) for three
different values of $s^2$} \centering
\begin{tabular}{c c c c c}
\hline\hline $s^2$       &\qquad 0.5       &\qquad 1       &\qquad
2 &\qquad 3\\ [0.5ex] $\bar s^2$       &\qquad 0.2372 &\qquad
0.4744 &\qquad 0.9488 &\qquad 1.4232\\ [0.5ex] \hline
 $F[1/s^2]$       &\qquad 0.0140     &\qquad  0.0667  &\qquad  0.2735  &\qquad  0.5834 \\
 $H[1/s^2]$       &\qquad 0.0104     &\qquad  0.0120  &\qquad  0.0119  &\qquad  0.0110 \\
 $L[1/s^2]$       &\qquad 0.0089      &\qquad  0.0094  &\qquad  0.0078 &\qquad  0.0064 \\ [1ex]
 \hline
 \end{tabular}
 \end{table}

\noindent The root mean square radius $\bar R$ in now given by
$\bar R^2=12R^2$, so a more meaningful reference parameter is
$\bar s\equiv r_c/\bar R=x_0/\sqrt{12}\cdot \lambda/R$. As a
function of $\bar R$, $\bar s$ and of the correlation strength
$c$, the expressions of the effective cross section and of the
parameter $\tau$ are

\begin{eqnarray}
{1\over\sigma_{eff}}={3\over7\pi \bar R^2}\Bigl\{1+c\cdot14\eta
F\Bigl({1\over12}{x_0^2\over \bar
s^2}\Bigr)+c^2\cdot{14\over15}\eta^2\Bigl({1\over12}{x_0^2\over
\bar s^2}\Bigr)\Bigr\}
\end{eqnarray}

\noindent and

\begin{eqnarray}
{1\over\tau\sigma_{eff}^2}\simeq{0,27\over\pi^2 \bar
R^4}\Bigl\{1+c\cdot400\eta H\Bigr({1\over12}{x_0^2\over \bar
s^2}\Bigl)+c^2\cdot\eta^2\Bigl[{x_0^2\over\bar
s^2}{0,64\over3}+800L\Bigr({1\over12}{x_0^2\over \bar
s^2}\Bigl)\Bigr]\Bigr\}
\end{eqnarray}

\subsection{Superposition of Poissonians}

A particular case, where all n-body correlations are important and
which can be worked out explicitly, is when the parton
distribution is given by the superposition of different
Poissonians. The superposition of Poissonians is naturally
obtained when introducing diffraction in a multi-channel eikonal
model of high energy hadronic
interactions\cite{Treleani:2007gi}\cite{Treleani:2008hq}\cite{Frankfurt:2008vi}.
To have some indication on this case we consider the simplest
possibility where the probability $P_n$, to find $n$ partons
within a given kinematical range, is given by the sum of two
Poissonians with average numbers $n_1$ and $n_2$

\begin{eqnarray}
P_n&=&\Bigl[\gamma{n_1^n\over n!}\esp{-n_1} +(1-\gamma){n_2^n\over
n!}\esp{-n_2}\Bigr]\nonumber\\ n_{1,2}&=&\int
n_{1,2}(x,b)dxd^2b,\qquad n_{1,2}(x,b)=G(x)g(b,R_{1,2}^2)
\end{eqnarray}

\noindent here $\gamma$ gives the relative weight of the two
Poissonians and the integration limits in $x$ are defined by the
kinematical range relevant to the case of interest. Notice that,
as we want to disentangle the effect of correlations in the
transverse coordinates, while $n_{1}(x,b)\neq n_{2}(x,b)$, the
integrated values  $n_1$ and $n_2$ are equal. The average density
of partons with fractional momentum $x$ and transverse coordinate
$b$ is

\begin{eqnarray}
\langle n\rangle&=&\sum_{n=1}^{\infty}nP_n=\bigl[\gamma n_1+(1-\gamma)n_2\bigr]= \int D(x,b)dxd^2b\quad;\nonumber\\
D(x,b)&\equiv&G(x)\bigl[\gamma g(b,R_1^2)
+(1-\gamma)g(b,R_2^2)\bigr]
\end{eqnarray}

\noindent while for the average density of pairs of partons with coordinates $x_1,b_1$ and $x_2,b_2$ one obtains

\begin{eqnarray}
\langle n(n-1)\rangle&=&\sum_{n=1}^{\infty}n(n-1)P_n=\bigl[\gamma
n_1^2+(1-\gamma)n_2^2\bigr]\nonumber\\&&\qquad\qquad\qquad\quad=\int
D_2(x_1,b_1;x_2,b_2)dx_1dx_2d^2b_1d^2b_2\quad;\nonumber\\
D_2(x_1,b_1;x_2,b_2)&\equiv&G(x_1)G(x_2)\bigl[\gamma
g(b_1,R_1^2)g(b_2,R_1^2)+(1-\gamma)g(b_1,R_2^2)g(b_2,R_2^2)\bigr]
\end{eqnarray}

\noindent and analogously

\begin{eqnarray}
D_n(x_1,b_1\dots x_n,b_n)&\equiv&G(x_1)\dots G(x_n)\nonumber\\&&\quad\times\bigl[\gamma g(b_1,R_1^2)\dots g(b_n,R_1^2)+(1-\gamma)g(b_1,R_2^2)\dots g(b_n,R_2^2)\bigr]
\end{eqnarray}

\noindent The expression of the inclusive cross section of $N$ independent parton collisions $\sigma_N$ hece is

\begin{eqnarray}
\sigma_N={1\over N!}\sigma_S^N\times\Bigl\{\gamma^2 \int \bigl[g(\beta,2R_1^2)\bigr]^Nd^2\beta&+&2\gamma(1-\gamma)\int \bigl[g(\beta,R_1^2+R_2^2)\bigr]^Nd^2\beta\nonumber\\&&\qquad+(1-\gamma)^2\int \bigl[g(\beta,2R_2^2)\bigr]^Nd^2\beta \Bigr\}
\end{eqnarray}

\noindent The actual calculations are carried out for the case of
the Gaussian parton density, where the effective cross section and
the parameter $\tau$ of the triple scattering inclusive cross
section are given by

\begin{eqnarray}
{1\over\sigma_{eff}}={3\over4\pi\bar
R^2}\Bigl\{\gamma^2\cdot{1\over2\alpha}+2\gamma(1-\gamma)\cdot{1\over2}+(1-\gamma)^2\cdot{1\over2(2-\alpha)}\Bigr\}
\end{eqnarray}

\noindent and

\begin{eqnarray}
{1\over\tau\sigma_{eff}^2}={3\over4\pi\bar
R^4}\Bigl\{\gamma^2\cdot\Bigl({1\over2\alpha}\Bigr)^2+2\gamma(1-\gamma)\cdot\Bigl({1\over2}\Bigr)^2+(1-\gamma)^2\cdot\Bigl({1\over2(2-\alpha)}\Bigr)^2\Bigr\}
\end{eqnarray}

\noindent where we made the positions

\begin{eqnarray}
R_1^2=\alpha R^2,\quad R_2^2=(2-\alpha) R^2\quad{\rm and}\quad R^2={3\over2}\bar R^2\nonumber
\end{eqnarray}

\noindent with $\bar R^2$ the mean square hadron radius measured in the generalized parton distributions.

One recognizes that the three different terms in the curly brackets are the contributions to the scale factors due to all possible combinations of the different sizes of the two interacting hadrons ($R_1$-$R_1$, $R_1$-$R_2$+$R_2$-$R_1$ and $R_2$-$R_2$)\cite{Treleani:2008hq}.

To make contact with the general formalism previously discussed, one may identify the correlation term by the relation

\begin{eqnarray}
D_2(x_1,b_1;x_2,b_2)=\bigl[D(x_1,b_1)D(x_2,b_2)+C(x_1,b_1;x_2,b_2)\bigr]
\end{eqnarray}

\noindent which gives

\begin{eqnarray}
C(x_1,b_1;x_2,b_2)&=&G(x_1)G(x_2)\Bigl\{\bigl[\gamma g(b_1,R_1^2)g(b_2,R_1^2)+(1-\gamma)g(b_1,R_2^2)g(b_2,R_2^2)\bigr]\nonumber\\&&\qquad\qquad\qquad\qquad\quad-\bigl[\gamma g(b_1,R_1^2)
+(1-\gamma)g(b_1,R_2^2)\bigr]\nonumber\\&&\qquad\qquad\qquad\qquad\qquad\quad\times\bigl[\gamma g(b_2,R_1^2)
+(1-\gamma)g(b_2,R_2^2)\bigr]\Bigr\}\nonumber\\
&=&\gamma(1-\gamma)G(x_1)G(x_2)\Bigl\{\bigl[g(b_1,R_1^2)-g(b_1,R_2^2)\bigr]\nonumber\\&&\qquad\qquad\qquad\qquad\qquad\quad\times\bigl[g(b_2,R_1^2)
-g(b_2,R_2^2)\bigr]\Bigr\}\end{eqnarray}

\noindent The correlation strength $c$ is hence expressed as a function of the relative weight of the two Poissoninas $\gamma$ by the relation

\begin{eqnarray}
c=\gamma(1-\gamma)
\end{eqnarray}

\noindent while the correlation length $r_c$, now defined by the change of sign of the correlation terms for $b_1$ or $b_2=r_c$, is given here below as a function of the mean square root hadron radius $\bar R$ and of the parameter $\alpha$, which controls the relative value of the two transverse radii $R_1$ and $R_2$:

\begin{eqnarray}
r_c=\bar R\Bigl[{1\over3}\alpha(2-\alpha){\rm ln}\Bigl({2-\alpha\over\alpha}\Bigr)\Bigr]^{1\over2}
\end{eqnarray}

\noindent The case discussed in \cite{Frankfurt:2008vi}
corresponds to $c={1\over4}$ and $r_c={\bar R\over2}\sqrt{\rm
ln}3\simeq.52\bar R$. With $\bar
R2=.42$fm$^2$\cite{Frankfurt:2003td} one obtains
$\sigma_{eff}\simeq30$mb, too large to explain the value of
$\sigma_{eff}$ observed by CDF\cite{Abe:1997bp}\cite{Abe:1997xk}.

Notice that in the two-channel eikonal model discussed
in\cite{Treleani:2007gi} one obtains a value of $\sigma_{eff}$ in
agreement with the experimental indication. The reason is that, in
the model, compact hadronic configurations are characterized by a
stronger Pomeron coupling, which corresponds associating a higher
partonic population to the compact configurations. In the present
case the distribution in the number of partons is, on the
contrary, the same in the two configurations with transverse
distances $R_1$ and $R_2$ (namely $n_1=n_2$, after integrating on
$b$). One may hence conclude that the experiment indicates that
the fluctuation of the whole hadron structure in its transverse
size, to the extent suggested by diffraction, is not enough to
explain the value of $\sigma_{eff}$, which may, on the contrary,
require the introduction of correlation terms of the kind
discussed here above in subsections 3.1 and 3.2, or/and of
correlations in fractional momenta. In this last instance the
multiparton distribution of the hadron structure is different from
a Poissonian, also after integrating on the parton's kinematical
variables.

\section{Concluding remarks}

Multiple parton interactions are going to play an important role at the LHC, both in the description of the properties of the minimum bias and of the underlying event and as a background to various channels of interest for the search of new physics. The study of MPI represents moreover the basic handle to obtain information on unknown non-perturbative features of the hadron structure, namely the correlations between partons.

In the present paper we have tried to identify the quantities which are most suitable to obtain the information on the non-perturbative features associated to the presence of MPI, in the case where the MPI are dominated by independent collisions initiated by different pairs of partons. To this purpose, in addition to the inclusive cross sections considered till now in hard processes, we made use also of the information provided by the "exclusive" cross sections.
Following a previous article where the "exclusive" cross sections were introduced\cite{Calucci:2008jw}, we have hence analyzed MPI considering systematically all terms up to triple scatterings. The inclusive and the exclusive MPI cross sections are linked by the sum rules of Eq.(2) and, by checking the number of terms needed to saturate the sum rules, one has a direct control on the importance of the unitarity corrections. The case where no more that three parton collisions give significant contributions leads to very simple relations between inclusive and "exclusive" cross sections, the relations (18) and (20) which, being a consequence of the definitions of the cross sections, hold rather in general. One may thus obtain the values of the relevant parameters, $\sigma_{eff}$ for the double parton collisions and the dimensionless parameter $\tau$ for the triple (having defined in that case the scale factor as $\tau\sigma_{eff}^2$).

A convenient way to measure the scale factors may be through Eq.s(21), which make use only of single and double collisions terms. Notice that if the sum rules of Eq.(2) are saturated with three terms in a given phase space window (and hence in a given interval of $x$ values) Eq.s(21) must hold. By measuring $\sigma_S$, $\tilde\sigma_1$ and $\tilde\sigma_2$ as a function of $x$ in the given interval, one may hence obtain a reliable information on the dependence of the correlation terms on $x$.

In our approach correlations are introduced in the most general way, as deviations of the multi-parton distributions from the Poissonian. Rather than trying to propose definite correlation models, our philosophy is hence to identify the observable quantities which are most suitable to obtain information on the correlation terms of the hadron structure. To have an idea of where the correlation parameters (correlation length and correlation strength) may be most relevant we have considered a few simplest cases. Of course correlations will depend on all variables and in particular on fractional momenta, because of conservation laws. Nevertheless conservation laws may not play a very important role when the parton population is large, namely at small $x$. In Section 3 he have worked out in full detail three simplified instances, where the dependence of correlations on $x$ may be neglected and which allows a full (or almost full) analytic treatment (Gaussian and exponential parton densities and correlations, multi-parton distribtuions given by a superposition of Poissonian). The relevant non perturbative information, namely the quantities $\sigma_{eff}$ and $\tau$, are given as a function of the correlation parameters in Eq.s (26, 27), Eq.s. (30, 31) and Eq.s (37, 38) in the three cases. Notice that the hypothesis of a negligible dependence of correlations on fractional momenta is easily tested experimentally by looking at the dependence of $\sigma_{eff}$ and $\tau$ on $x$.

In our discussion we did not allow for the differences between
partons (gluons and quarks, different flavors, valence and sea).
When considering definite reaction channels, the relations
obtained have hence to be adapted, taking into account that the
information on correlations will have to be related to the
different kinds of initial state partons involved in the
interactions.

\begin{center}
    {\bf APPENDIX}
 \end{center}

When looking at the effects of the correlation terms of the original multi-parton distribution in processes where three or more partons undergo hard scattering, a natural question is how to deal with higher order correlation terms. In this appendix we sketch the procedure to deal with this problem. When taking in consideration three-body correlations the starting point is:
\begin{eqnarray}
{\cal Z}[J+1]&\equiv&{\rm exp}\biggl\{\int D(u)J(u)du+{1\over 2}\int C(u,v)J(u)J(v)dudv\nonumber\\
&&\qquad\qquad\qquad\qquad\quad+{1\over 6}\int T(u,v,w)J(u)J(v)J(w)dudvdw\biggl\}\cr
&=&\sum_{n}{1\over n!}\int J(u_1)\dots J(u_n)D_n(u_1\dots u_n)du_1\dots du_n\nonumber
\end{eqnarray}
 The possibility of giving a close form to $\cal Z$ does not exist any more, but a general procedure is available, and well known \cite{Hori}: every functional admitting a formal series expansion can be expressed as
 $$\Phi [J]=\Phi[\delta/\delta\chi]\exp\int J(u)\chi(u)du\Big|_{\chi=0}\;;$$    so the previous expression may be rewritten in the following form:
\begin{eqnarray}
{\cal Z}[J+1]&\equiv&{\rm exp}\Big[{1\over 6}\int T(u,v,w)\frac {\delta}{\delta(u)}\frac {\delta}{\delta(v)}\frac {\delta}{\delta(w)}dudvdw \Big]\cr
&&\times{\rm exp}\biggl\{\int [D(u)+\chi (u)]J(u)du+{1\over 2}\int C(u,v)J(u)J(v)dudv
\biggl\}\Big|_{\chi=0}\;.\nonumber
\end{eqnarray}
The second exponential will give rise to terms similar to the ones in Eq. (6,7), the only difference is in the terms $a_n$, which contain $D+\chi$ instead of $D$.
The whole expression is suitable for an expansion in $T$. When acting with the derivatives one finds that every term $T$ is connected either with three terms $M(u,v)$, defined by the relation below, ending in turn on a density $D$ or with two of them, one of which ends on one $D$ and the other is closed on the same term $T$.

\begin{eqnarray}a_n=(-1)^{n+1}\int&& D_A(u_1)\hat{\sigma}(u_1,u_1')C_B(u_1',u_2')
       \hat{\sigma}(u_2',u_2)
       C_A(u_2,u_3)\dots\nonumber\\
       &&\quad\qquad\qquad\qquad\qquad\qquad\cdots
       \hat{\sigma}(u_n,u_n')D_B(u_n')
  \prod_{i=1}^ndu_idu_i'\nonumber\\
\equiv(-1)^{n+1}\int&& D_A(u_1)\hat{\sigma}(u_1,u_1')M_n(u_1',u_n')\hat{\sigma}(u_n,u_n')D_B(u_n')
  \prod_{i=1}^ndu_idu_i'\nonumber
\end{eqnarray}

\noindent In particular the first two terms are:
\begin{eqnarray}
&&\int T(u,v,w)\hat{\sigma}(u,u')D(u')\hat{\sigma}(v,v')D(v')\hat{\sigma}(w,w')D(w')dudvdwdu'dv'dw'\nonumber\\{\rm and}
&&\int T(u,v,w)\hat{\sigma}(u,u')\hat{\sigma}(v,v')C(u',v')\hat{\sigma}(w,w')D(w')dudvdwdu'dv'dw'
\;;\nonumber
\end{eqnarray}

\noindent and one expects the first to be the most important.

\end{document}